\pdfoutput=1

\documentclass[aps,prb,reprint,showpacs,superscriptaddress]{revtex4-1}
\usepackage{graphicx}
\usepackage{graphics}
\usepackage{subfigure}
\usepackage{amsmath}
\usepackage{amssymb}
\usepackage{amsfonts}
\usepackage{dcolumn}
\usepackage{dsfont}
\usepackage{epstopdf}
\usepackage{latexsym}
\usepackage{rotating}
\usepackage{color}
\usepackage{latexsym}
\usepackage{bbm}
\usepackage{float}
\usepackage{epsfig}
\usepackage{epsf}
\usepackage{psfrag}
\usepackage{bm}
\usepackage{amsthm}
\usepackage{eucal}
\usepackage{mathrsfs}
\usepackage{url}
\usepackage{braket}
\usepackage[none]{hyphenat}
\usepackage{color} % use colored text in latex

\usepackage{hyperref}
\hypersetup{
colorlinks=true,final=true,
        linkcolor=blue,
        citecolor=blue,
        filecolor=blue,
        urlcolor=blue,
}
\begin{document}
\title{Interplay of magnetism and band topology in Eu$_{1-x}$Ca$_x$Mg$_2$Bi$_2$ (x=0, 0.5) from first principles study}

\author{Amarjyoti Choudhury}
\author{N. Mohanta}
\author{T. Maitra}
\email{tulika.maitra@ph.iitr.ac.in}
\affiliation{Department of Physics, Indian Institute of Technology Roorkee, Roorkee - 247667, Uttarakhand, India}
\date{\today}

\begin{abstract}
Recent discovery of the time reversal symmetry breaking magnetic Weyl semimetals has created a huge surge of activities in the field of quantum topological materials. In this work, we have studied systematically the ground state magnetic order, electronic structure and the interplay between the magnetic order and band topology in one such materials, EuMg$_2$Bi$_2$ (EMB) and its Ca doped variant using first principles method within the framework of density functional theory (DFT). The detailed investigation unravels the existence of different topological phases in this single material which can be tuned by an external probe such as magnetic field or chemical substitution. Our DFT calculations including Coulomb correlation (U) and spin-orbit (SO) interaction within GGA+U+SO approximation confirms that the magnetic ground state of EMB is A-type Antiferromagnetic (A-AFM) with Eu magnetic moments aligned along the crystallographic $a$ or $b$ direction. Although the ground state of EMB is A-AFM, the Ferromagnetic (FM) state lies very close in energy. We observe a single pair of Weyl points connecting valence and conduction band very close to the Fermi level (FL) along $\Gamma$-A direction in the FM state of EuMg$_2$Bi$_2$ with Eu moments aligned along crystallographic $c$ direction. On doping 50\% Ca at Eu sites, we observe single pair of Weyl points moving closer to the FL which is highly desirable for application purposes. Further we observe that the separation between the Weyl points in the pair decreases in doped compound compared to that in the parent compound which has direct consequence on anomalous Hall conductivity (AHC). Our first principles calculation of AHC shows high peak values exactly at these Weyl points and the peak height decreases when we dope the system with Ca. Therefore, Ca doping can be a good external handle to tune AHC in this system.     
\end{abstract}

\maketitle

\section{Introduction} 
Systems harbouring magnetic order driven topological phases have recently drawn a great deal of attention from the researchers in condensed matter physics and materials science due to their versatility and tunability in potential device applications such as spintronics\cite{stp2,stp3,stp4,WSMS}. Since the discovery of topological insulators \cite{TI}, the field of topological phases of matter has exploded with activities leading to the discovery of many exotic topological states in materials such as Dirac semimetal\cite{DSMS,DSMS1}, Weyl semimetal\cite{WSMS}, nodal line semimetal\cite{NSMS,NSMS1} etc. The robustness of symmetry protected topological states in these systems has immense implications for device applications. Very recently magnetic materials hosting topological states with strong correlation between magnetism and topology have come to fore. In topological semimetals a fourfold degenerate Dirac point can appear at the band crossing of two bands each of which is doubly degenerate due to Kramer's degeneracy when both inversion symmetry (P) and time reversal symmetry (T) are present. When either inversion symmetry (P) or time reversal symmetry (T) is broken, the bands become non-degenerate leading to a two-fold degenerate point at the band crossing called Weyl point. Thus a single Dirac point can break into two Weyl points with opposite chirality. Many nonmagnetic Weyl semimetal with broken inversion symmetry have been found till date (e.g. in TaAs \cite{B.Q,Huang,Sankar} and WTe$_2$\cite{Arita} family of compounds) whereas magnetic Weyl semimetals with broken time reversal symmetry are still rare.      

%%%%%%%%%%%%%%%%%%%%%%%%%%%%%%%%%%%%%%%%  
       \begin{figure}
	%\vspace{-1.0cm}
	\begin{center}
		\includegraphics[width=7.0cm]{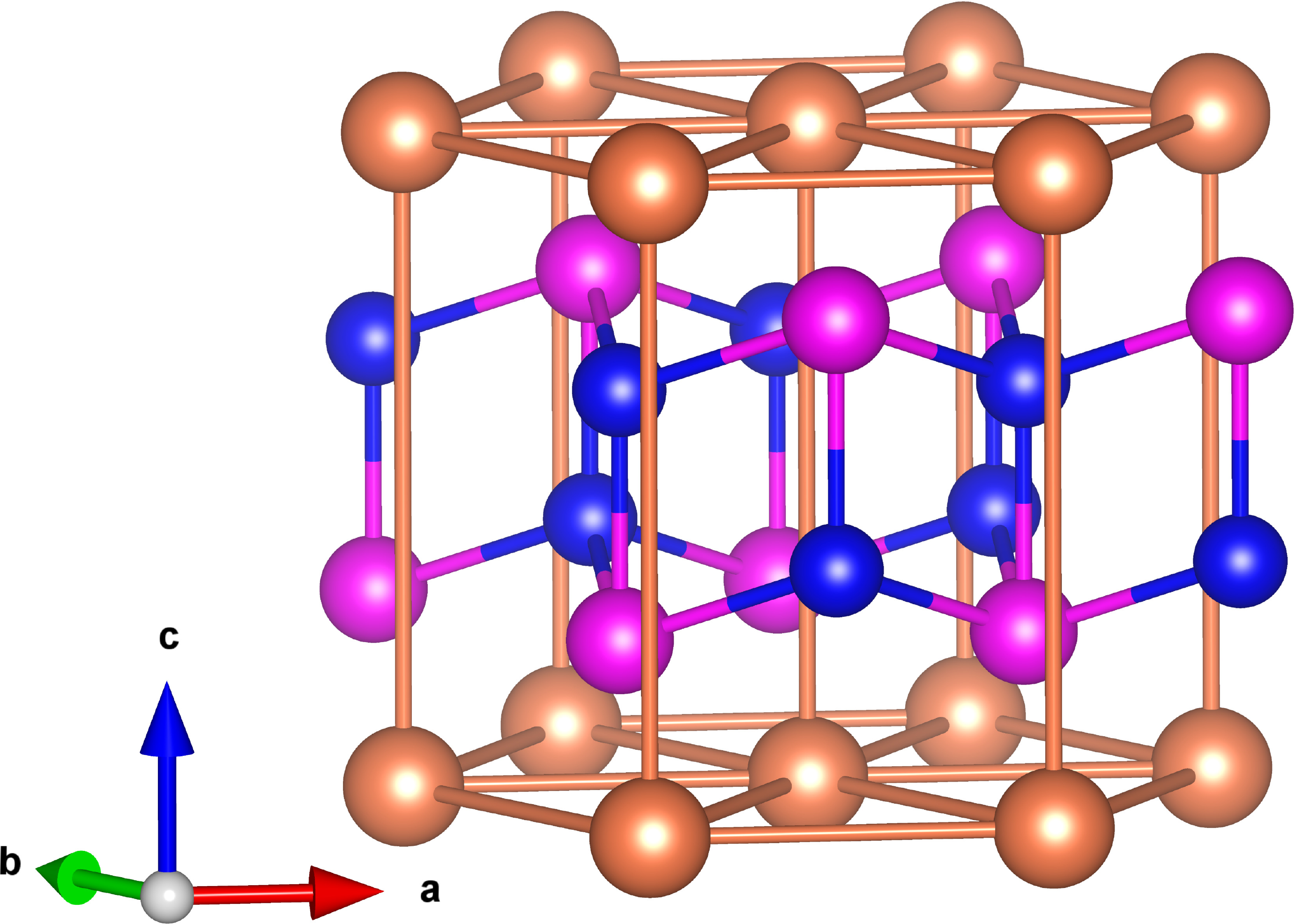}
		\caption{\label{Fig1} \noindent The Hexagonal crystal structure of EuMg$_2$Bi$_2$ with space group P$\bar{m}$31 where Eu, Mg and Bi ions are shown in brown, blue and pink colours respectively.}
			\end{center}
           \end{figure}
%%%%%%%%%%%%%%%%%%%%%%%%%%%%%%%%%%%%%%% 
In non-magnetic materials, the Weyl points (WPs), which appear due to breaking of inversion symmetry, exist always as multiple of four in the full Brillouin Zone (BZ) whereas in case of magnetic materials where the inversion symmetry is preserved but time reversal symmetry is broken due to intrinsic magnetism, the possibility of finding a single pair of Weyl points exists\cite{2WP}. The presence of WPs in a material can give rise to remarkable emergent phenomena such as the chiral anomaly effect, large magnetoresistance (MR), strong intrinsic anomalous Hall and spin Hall effects etc.\cite{felser} The presence of a single pair of Weyl points in a material is highly desirable for applications of such effects, for example, the anomalous Hall conductivity (AHC) depends on the separation between the partners in a single pair of Weyl points\cite{AHC}. Further, a magnetic material which hosts topological states where different long range magnetic orders have competing energies can be extremely useful because of their tunability by external magnetic field. The change in magnetic configuration can substantially change the symmetry of the material, which may lead to a different topological state.  

To find a material with minimum number of Weyl points, i.e., a single pair of WPs in the entire BZ, is a challenging task. The presence of a single pair of WPs was predicted theoretically in the ferromagnetic (FM) phase of the MnBi$_2$Te$_4$\cite{MnBiTe}. However, experimental verification is not reported yet. Very recently, in EuCd$_2$As$_2$, both experimental (ARPES) and theoretical (DFT) work indicated the presence of a single pair of Weyl points in its FM state with magnetic moments of the Eu aligned along the $c$ axis \cite{AA,soh}. It has been also found from density functional theory calculations\cite{jyoti} that in EuCd$_2$As$_2$ the A-type AFM (ground state) and FM (excited state) are very close by in energy which was later supported by the experiment where it was shown that a tiny external magnetic field as small as 2T can turn this system FM from A-type AFM\cite{soh}. Therefore, EuCd$_2$As$_2$ is the only known material so far to host a single pair of Weyl points (an ideal Weyl semimetal)\cite{felser}. As discussed by Wang et al.\cite{AA}, the materials which are either AFM Dirac semimetal or AFM topological insulator with a tiny band gap provide a fertile ground to search for a single pair of WPs in their FM phase. Therefore, EuMg$_2$Bi$_2$ which is a topological insulator having a very tiny gap as observed by Marshall et al. \cite{Marshall} is an ideal compound to explore for finding a single pair of Weyl points which is the main objective of our study here. 

%%%%%%%%%%%%%%%%%%%%%%%%%%%%%%%%%%%%%%%%  
	\begin{figure}
	%\vspace{-1.0cm}
	\begin{center}
		\includegraphics[width=8.0cm]{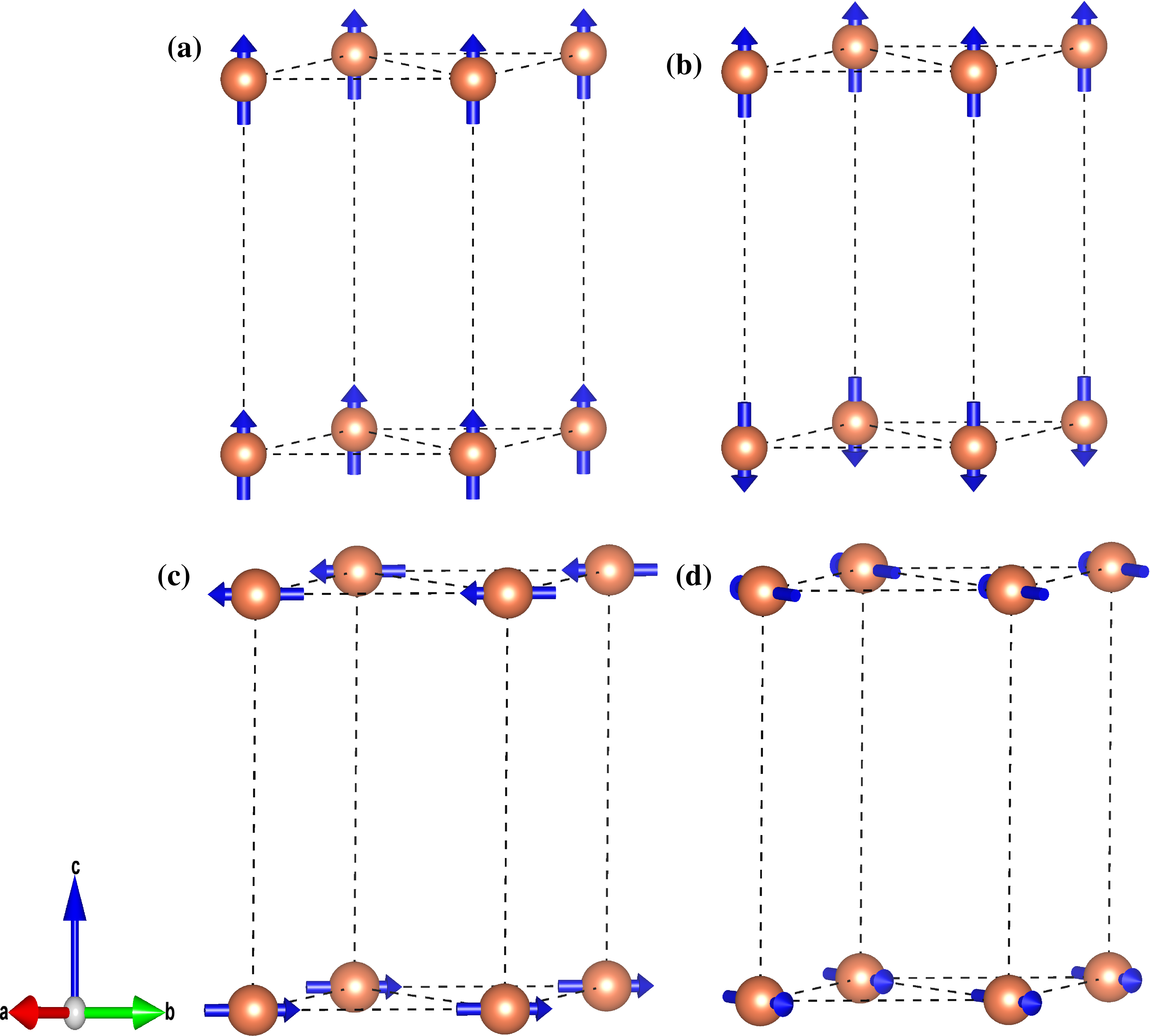}
	\end{center}
		\caption { \noindent Different magnetic configurations considered in our calculation (see text): (a) FMc, (b) A-AFMc, (c) A-AFMb (d) A-AFMx. }
		\label{Fig2}
	
\end{figure}
%%%%%%%%%%%%%%%%%%%%%%%%%%%%%%%%%%%%%%%

In the family of Zintl phase there are many layered 122-type BX$_2$Y$_2$ ternary intermetallic compounds which crystallize mostly into ThCr$_2$Si$_2$-type (tetragonal) structure, and rarely into CaAl$_2$Si$_2$-type (trigonal) structure\cite{H.Zhang, C.Zheng}. EuMg$_2$Bi$_2$ (EMB) crystallizes in the CaAl$_2$Si$_2$–type (trigonal) crystal structure with the space group P$\bar{3}$m1(164). The structure consists of rare earth magnetic Eu ions forming a triangular lattice in $ab$ plane with simple hexagonal stacking along $c$ axis separated by MgBi layers (see Fig.~\ref{Fig1}). Experiments confirm that the Eu$^{+2}$ ions with spin 7/2 in EuMg$_2$Bi$_2$ undergo a magnetic transition from the paramagnetic to antiferromagnetic state at a temperature close to 7K \cite{A.F.may}. The temperature dependent magnetic susceptibility measurement by May et al.\cite{A.F.may} indicated an anisotropic behaviour with susceptibility along $c$ being lower than that in the $ab$-plane from where the authors concluded that the magnetic moments are aligned along $c$ axis. Later, Pakhira et al. \cite{Pakhira} by analyzing the anisotropic magnetic susceptibility using the molecular field theory (MFT) proposed that the magnetic structure of the EMB to be a $c$-axis helix AFM where the magnetic moments of Eu are ferromagnetically aligned in $ab$ plane with a turn angle of 120$^\circ$ between adjacent Eu layers along the $c$ direction. Very recent neutron diffraction measurements in single crystal EuMg$_2$Bi$_2$ by two groups\cite{Marshall, pakhira1} reveal that the magnetic structure is A-type AFM (magnetic moments are ferromagnetically aligned in the $ab$ plane whereas they are antiferromagnetically aligned along $c$ direction) with Eu moments in the residing in the $ab$-plane. DFT calculations performed by Marshall et al. further shows that this compound is a topological insulator \cite{Marshall}. Marshall et al.\cite{Marshall1} have also reported in another very recent experimental work the Ca doping at Eu sites in EMB and observed that upon increasing the Ca doping the ground state magnetic structure changes from A-AFM to FM in this system. To understand the above all experimental observations which indicate a strong interplay of magnetic and topological properties in this compound similar to the compounds EuCd$_2$As$_2$ \cite{jyoti, AA}  and EuCd$_2$Sb$_2$ \cite{AAA} where topological properties are seen to vary with the magnetic configuration we have carried out a systematic study of magnetic order, electronic structure and topological properties of EMB and its Ca doped variant density functional theory calculations.

The paper is arranged in the following way. We provide the details of the methods used in our calculations in section II. In section III, we present our results in three subsections: (A) Magnetic order and electronic structure, (B) Magnetic order and Topological properties and (C) Ca doped EuMg$_2$Bi$_2$: Topological properties). Section IV contains conclusions and section V acknowledgements.

\section{Methods}
\label{meth}
The total energy and electronic band structure calculations for the compound EuMg$_2$Bi$_2$ were performed using density functional theory (DFT) calculations with Perdew-Burke-Ernzerhof generalized gradient approximation (PBE-GGA)\cite{PBE} exchange correlation functional. Plane-wave basis set and projector-augmented wave\cite{PAW} method were used as implemented in the Vienna Ab-{\sl{initio}} Simulation Package (VASP)\cite{kresse, kresse2}. The structural parameters of EMB are taken from the experiment\cite{Marshall} and are then fully optimized. Both Coulomb correlation and spin orbit(SO) interaction are included in our calculation since the Eu has strongly localized 4f electrons. For self-consistent calculation, we used $\Gamma$-centered Monkhorst-Pack \cite{Mon} (11x11x7) {\bf k} point mesh in the Brillouin zone (BZ). The kinetic energy cut off for the plane wave basis set was set to 340 eV. GGA+U calculations are carried out within Dudarev formalism \cite{Duda} with an effective Hubbard interaction (U$_{eff}$=U-J where U is Coulomb correlation and J is Hund's exchange). Results are presented for U$_{eff}$=5eV applied to Eu 4f states though the value was varied between 3eV to 11eV to verify the robustness of our results. To study the topological properties of EuMg$_2$Bi$_2$ and Ca doped variants, we used the software packages Wannier90 \cite{Mostofi} in combination with WannierTools\cite{Wu} (WT). 
Wannier90 uses Maximally Localized Wannier Functions (MLWF)\cite{Vander} to obtain the tight binding model by fitting the DFT bands which is then used by the WT to calculate various topological and transport properties.  
We have identified the position of Weyl points in the Brillouin Zone and the corresponding Berry-curvature, Chern number etc. using WT. We also calculated the surface state ARPES spectrum using the iterative Green's function method as impletented in WT. AHC as a function of energy is also computed for both EuMg$_2$Bi$_2$ and the Ca doped cases using WT.

\section{Results and Discussion}
\label{res}
\subsection{Magnetic order and electronic structure}
In order to establish the ground state magnetic order of EuMg$_2$Bi$_2$, total energy calculations were performed for ferromagnetic (FM) as well as A-type antiferromagnetic (A-AFM) configurations within GGA, GGA+U and GGA+U+SO approximations using the optimized structural parameters. Further, we have also examined the configurations of FM and A-AFM where Eu moments point along crystallographic $c$ and $b$ directions as well as in a direction exactly in-between crystallographic $a$ and $b$ axes (we denote it by $x$). Some of these magnetic configurations are shown in Fig~\ref{Fig2}. In the Table~\ref{Tab1} we present the total energies calculated for A-AFM and FM by considering various approximations. One can clearly see that the total energies for FM and A-AFM are energetically quite close to each other within all the approximations considered. The FM state is found to be the lowest within GGA approximation. However, inclusion of Coulomb Correlation (U) within GGA+U approximation immediately makes the A-AFM to be the most energetically favorable. The A-AFM ground state remains robust as we vary U$_{eff}$=U-J (where U is the Coulomb correlation and J is Hund's exchange) between 3 to 11 eV. The difference of energies of A-AFM and FM as a function of U$_{eff}$ is listed in Table~\ref{Tab2}. 

%%%%%%%%%%%%%%%%%%%%%%%%%%%%%%%%%%%%%%%%%%%%%%%%
\begin{figure}
	%\vspace{-1.0cm}
	\begin{center}
		\includegraphics[width=6.0cm]{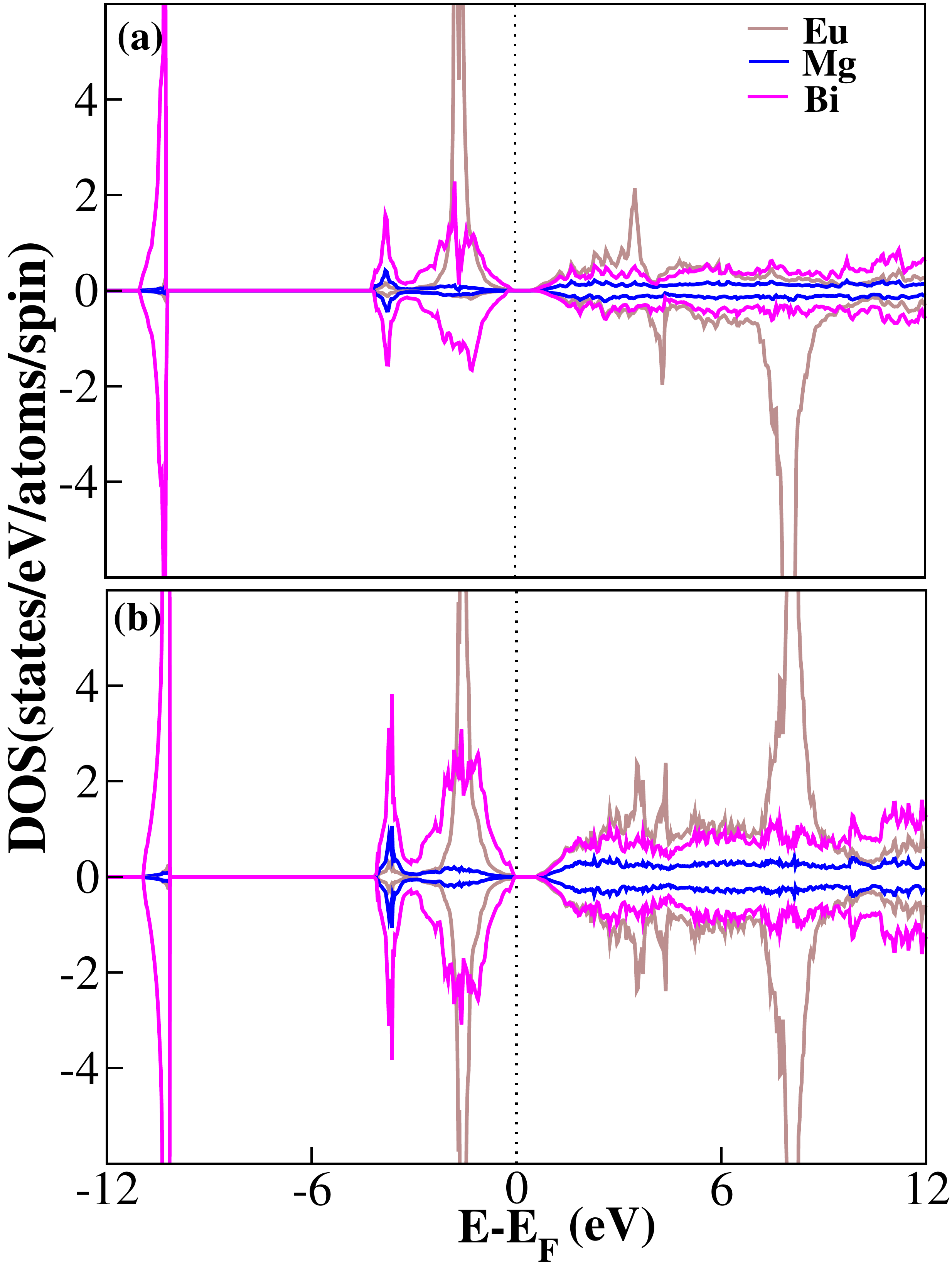}
		
		\caption{ \noindent Partial densities of states for (a) FM within GGA+U (b) A-AFM within GGA+U with
			U$_{eff}$ = U ${-}$J = 5 eV. Here the spin up and down states are shown on positive and negative y axis respectively.}
		     
		    \label{Fig3}
	\end{center}
\end{figure}
%%%%%%%%%%%%%%%%%%%%
\begin{table}
	\begin{center}
		\caption{Total energy (in meV) per formula unit computed with reference to the ground state for the FM and A-AFM magnetic configurations within various approximations. Here Eu moment direction is denoted by the $\mu$.}
		
		\label{Tab1}
			\begin{tabular}{m{10em} m{2cm} m{2cm}}
			\hline\hline	
			XC &    A-AFM  &      FM \\ 
			\hline
			GGA    &  0.202 &  0.00 \\
			GGA+U & & \\
			U$_{eff}$= 3eV & 0 & 1.096 \\
			U$_{eff}$ =5eV & 0  & 1.406\\
			GGA+U+SO & & \\
			U$_{eff}$=5eV & & \\
			$\mu \parallel b$ axis & 0.0 &1.4010 \\
			$\mu \parallel c$ axis &  0.2005 &1.3930 \\
			$\mu \parallel x$ axis  &0.0030 &1.4015 \\
			\hline
		\end{tabular}
		
	\end{center}
\end{table}
%%%%%%%%%%%%%%%%%%%%%%%%%%%%%%%%%%%%%

%%%%%%%%%%%%%%%%%%%%%%%%%%%%%%%%%%%%%%%%%%%%%
\begin{table}
	\begin{center}
		\caption{The variation of total energy difference between FM and A-AFM ($\Delta$E) with U$_{eff}$.} 
		\label{Tab2}
		\begin{tabular}	{m{6em} m{1.0cm} m{1.0cm} m{1.0cm} m{1.0cm} m{1.0cm}  } 
			\hline\hline
			%\hline
		     U$_{eff}$(eV)  & 0 &  3 & 5 & 7 & 11  \\
			\hline
			$\Delta$E(meV)	&  -0.202 &  1.096 & 1.406 & 1.613 & 1.909 \\
			\hline\hline
		\end{tabular}	
	\end{center}
\end{table}
%%%%%%%%%%%%%%%%%%%%%%%%%%%%%%%%%%

%%%%%%%%%%%%%%%%%%%%%%%%%%%%%%%%%%%%%%%%  
\begin{figure}
	%\vspace{-1.0cm}
	\begin{center}
		\includegraphics[width=9cm]{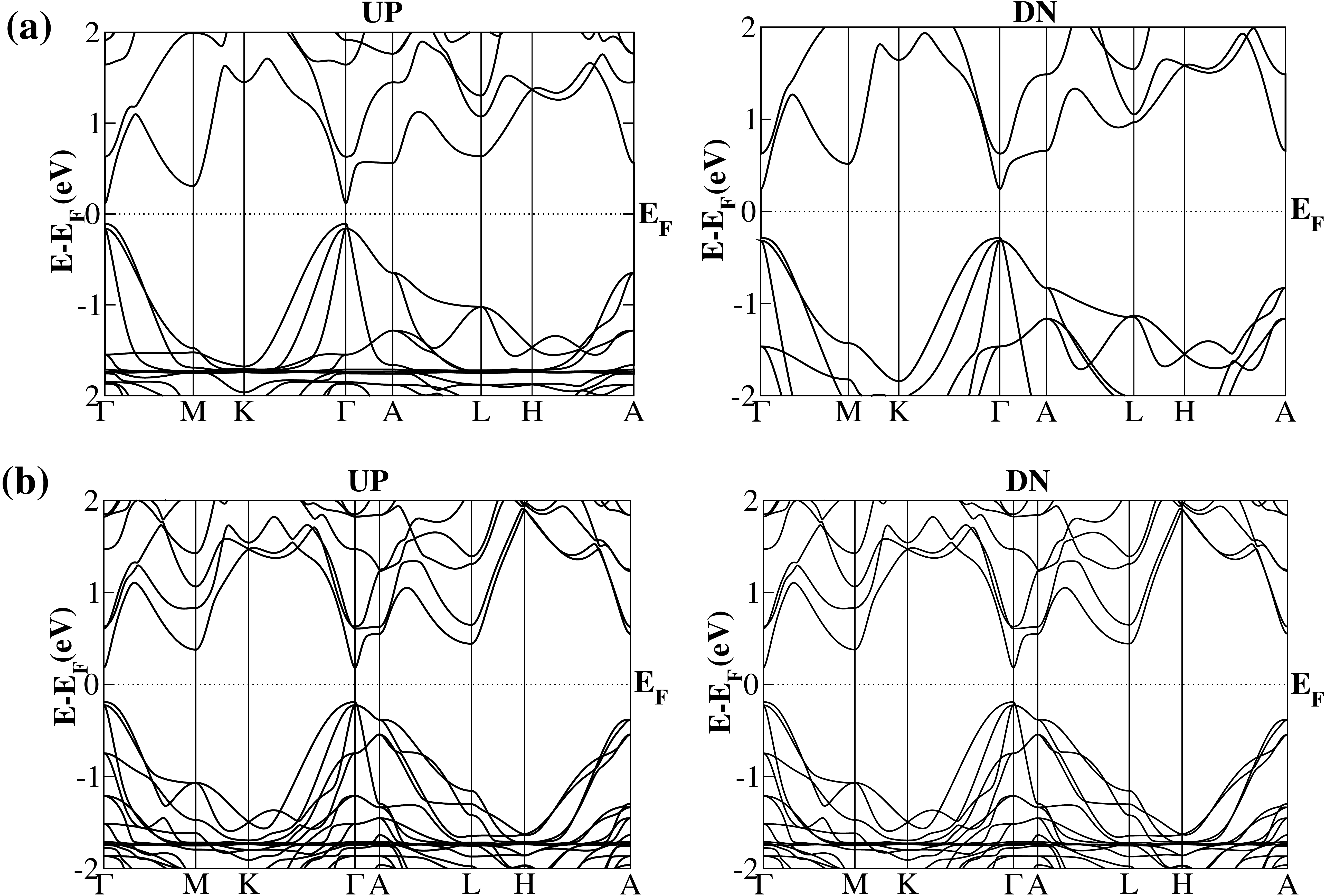}
		\caption{Band structure along high symmetry directions for (a) FM and (b) A-AFM calculated within GGA+U with U$_{eff}$=5eV.}
		\label{Fig4}
	\end{center}
\end{figure}
%%%%%%%%%%%%%%%%%%%%%%%%%%%%%%%%%%%%%%%

\begin{table}
	\begin{center}
		\caption{. The positions,energies and Chern numbers of pair of Weyl points in the BZ. (k$_x$,k$_y$,k$_z$ ) are positions in the reciprocal lattice of the unit cell} 
		\label{Tab3}
		\begin{tabular}	{m{1em} m{0.8cm} m{0.8cm} m{1.0cm} m{1.5cm} m{1.0cm} m{1.0cm}   }
			\hline\hline 
			 & & &EuMg$_2$Bi$_2$ \\
			\hline
			& &k$_x$& k$_y$&  k$_z$ & E-E$_f$ (meV)  & C  \\
			\hline
			&W$_1$ &0  & 0  & $+$0.02 &    0.081 & $+1$   \\
			%\hline
			&W$_2$ & 0 & 0  & $-$0.02 &   0.081 & $-1$ \\
			
			%&W$_3$ & 0.0013 & 0.0011	 & $+$0.0253  &  0.0366 & $+1$ \\
			\hline \\\\
			%&W$_4$ & 0.0013 & 0.0011  & $-$0.0253 &   0.0366 & $-1$ \
		 	  \hline\hline
		 	  && &Eu$_{0.5}$Ca$_{0.5}$Mg$_2$Bi$_2$\\
		 	  \hline
		 	  & &k$_x$& k$_y$&  k$_z$ & E-E$_f$ (meV)  & C  \\
		 	  \hline
		 	  &W$_1$ &0 &0 &$+$0.01 &  0.027 &$+1$\\
		 	  %\hline
		 	  &W$_2$ &0&0 &$-$0.01  &  0.027  &$-1$\\
		 	  \hline
		\end{tabular}	
	\end{center}
\end{table}

%&W$^{\mp}$$_1$ &k$_1$& k$_2$& k$_3$ & 2  \\

   %%%%%%%%%%%%%%%%%%%%%%%%%%%%%%%%%%%%%%%
\begin{figure}
	%\vspace{-1.0cm}
	\begin{center}
		\includegraphics[width=8.0cm]{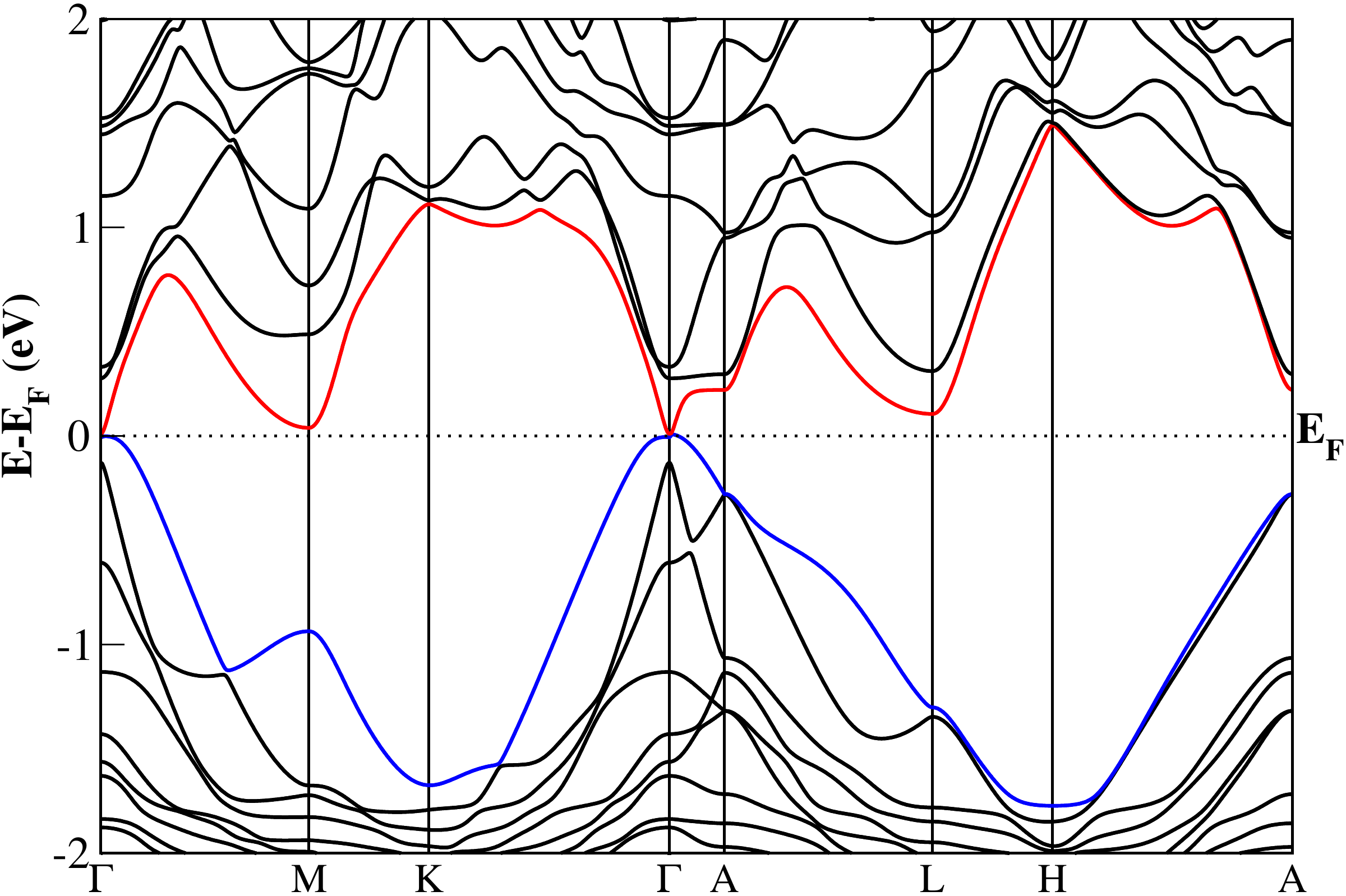}
		\caption{(a)Band structure of EuMg$_2$Bi$_2$ calculated within GGA+U+SO for A-AFM with Eu magnetic moments along crystallographic $b$ direction.}
		\label{Fig5}
	\end{center}
\end{figure}
%%%%%%%%%%%%%%%%%%%%%%%%%%%%%%%%%%%%%%%
%%%%%%%%%%%%%%%%%%%%%%%%%%
From Table~\ref{Tab1}, we further observe that the true magnetic ground state is A-AFM with Eu moments pointing along crystallographic $b$ (equivalently $a$ because of hexagonal symmetry) direction which is consistent with recent neutron diffraction measurements on this system where the authors observed Eu spin moments to lie in $ab$ plane\cite{Pakhira, Marshall}. However, Pakhira et al.\cite{Pakhira} weren't able to establish from their measurements whether the spins point along $b$ or along $x$ (as defined above). Our calculations clearly shows that the configuration with Eu spins pointed along $b$ (A-AFMb) has lower energy than that along $x$ (see Table~\ref{Tab1}). We further observe that the A-AFM and FM configurations with spins pointing along $c$ (denoted by A-AFMc and FMc respectively) as well as FM with spins pointing along $b$ (FMb) are very close by in energy. The smallness of energy differences implies that these excited states can be easily accessible via application of external magnetic field as has been seen in another isostructural compound EuCd$_2$As$_2$ experimentally\cite{ECAexp}. Our calculation of the magnetic anisotropy energy (MAE) (i.e., the energy difference ($\Delta$E) between the two magnetic configurations with spin moments lying in plane (easy axis) and spin moments lie out of plane (hard axis)) is very small ($\Delta$E $=$0.2 meV$/$Eu). The tiny MAE observed by us in combination with the weak inter-layer antiferromagnetic interaction revealed by a recent experimental study\cite{Marshall} imply that the magnetic state in EMB is vulnerable to the external magnetic field. Marshall et al., from their topological invariant Z$_2$ number calculations, found EuMg$_2$Bi$_2$ to be a topological insulator \cite{Marshall}. To investigate the correlation between magnetic order and topological properties in this compound as has been seen in other isostructural compounds EuCd$_2$As$_2$ and EuCd$_2$Sb$_2$ where topological properties depend on the magnetic configuration, we present our electronic structure results below. 

%%%%%%%%%%%%%%%%%%%%%%%%%%%%%%%%%%%%%%%

%
We have calculated the partial density of states (PDOS) for FM and A-AFM configurations within GGA+U approximation with U$_{eff}$ = 5 eV applied to Eu 4f states which we present in Fig. ~\ref{Fig3}. We observe insulating ground state for both the magnetic configurations. 
To get better understanding of the states closer to the FL, we have calculated the band structure along the high symmetry directions in the first Brillouin zone (BZ) for both spin up (UP) and down (DN) states for FM and A-AFM configurations as shown in Fig~\ref{Fig4}(a) and Fig~\ref{Fig4}(b) respectively. As can be seen from Fig ~\ref{Fig4}(a), in the FM configuration the band gap is less in the spin up channel as compared to the same along spin down channel indicating a strong exchange splitting in this system. In the A-AFM configuration the band gap is same in both the spin channels and has a value of about 0.4eV. The localized flat bands due to Eu 4f states can be seen close to 2eV
%%%%%%%%%%%%%%%%%%%%%%%%%%%%%%%%%%%%%%%
\begin{figure}
	%\vspace{-1.0cm}
	\begin{center}
		\includegraphics[width=8.0cm]{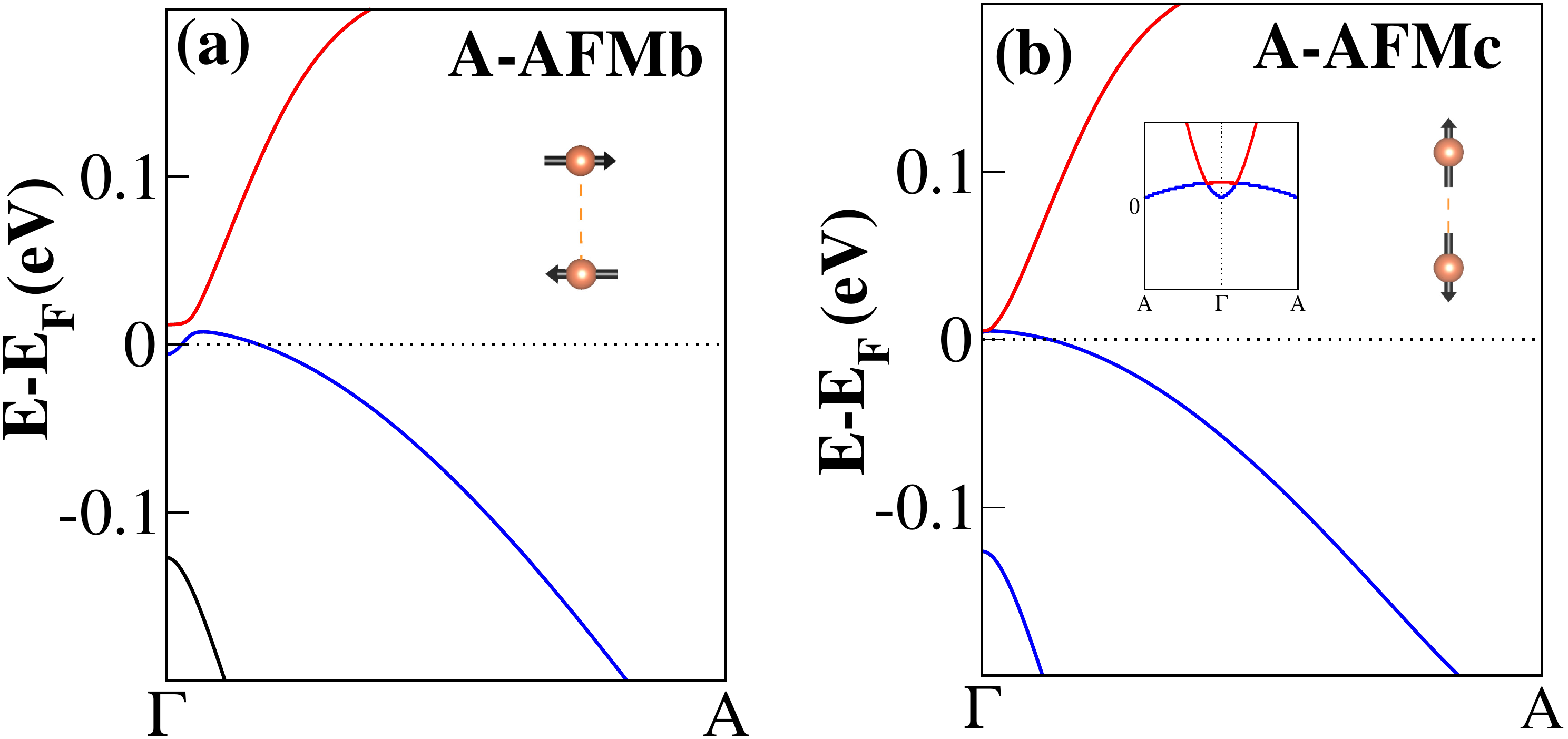}
		\caption{Bulk band structure of EMB along $\Gamma$-A direction for (a) A-AFMb (b) A-AFMc magnetic configurations.}
		\label{Fig6}
	\end{center}
\end{figure}
%%%%%%%%%%%%%%%%%%%%%%%%%%%%%%%%%%%%%%%
 below Fermi level which are further pushed down upon increasing $U$ value though the band gap remains same around 0.4 eV. As the true magnetic ground state of this compound is A-AFM with moments lying parallel to $b$ axis (A-AFMb) and also this compound contains heavy ions like Eu and Bi, we have calculated the band structure for A-AFMb magnetic configuration by including spin-orbit interaction in our calculation within GGA+U+SO approximation as shown is Fig~\ref{Fig5}. Comparing with the band structure obtained without SO (Fig~\ref{Fig4}(b)), one observes that the band gap around $\Gamma$ almost disappears in this case. We explore this region around $\Gamma$ point and related topological properties in detail and present our results in the next subsection. We also investigated the topological properties for various other magnetic orders which are close in energy to that of the ground state and discussed them in the next subsection. 
%%%%%%%%%%%%%%%%%%%%%%
\begin{figure}
	%\vspace{-1.0cm}
	\begin{center}
		\includegraphics[width=8.0cm]{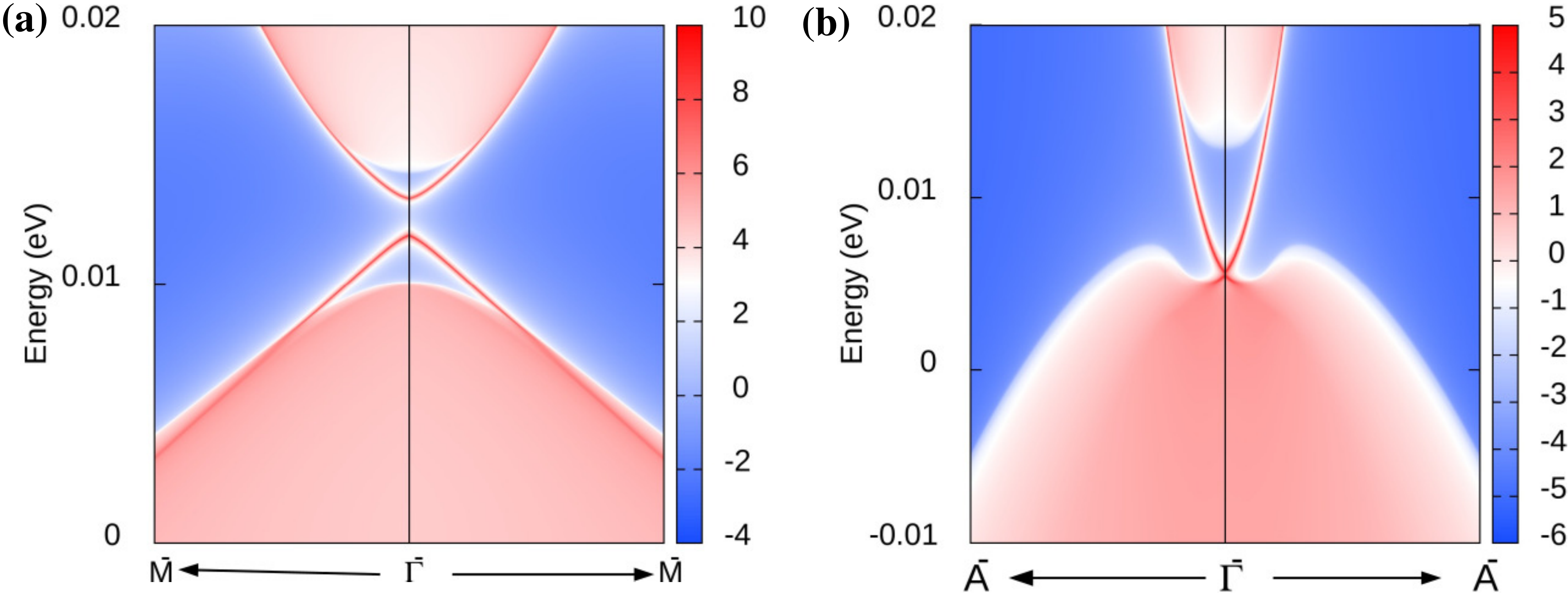}
		\caption{The surface states for EMB in A-AFMb magnetic configuration on (a) (001) and (b) (100) surface resepctively.}
		\label{Fig7}
	\end{center}
\end{figure} 
%%%%%%%%%%%%%%%%%%%%%%%%%%%%%%%%%%%%%%% 

\subsection{Magnetic order and Topological properties}

To investigate how topological properties of EMB depend on and vary with the underlying magnetic order, we have looked closely into the band structure of ground state magnetic order A-AFMb along with various other magnetically ordered states such as A-AFMc and FMc which are found to be energetically close to the ground state and hence can be tuned by external magnetic field. Looking at the band structure plot for A-AFMb configuration within GGA+U+SO presented in Fig~\ref{Fig5}, we see that except the region around $\Gamma$ point, particularly along $\Gamma$-A direction, there is a significant band gap between valence band and conduction band all along high symmetry path. In Fig~\ref{Fig6}(a) we present the zoomed in band structure plot close to the FL along $\Gamma$-A for AFMb configuration. One can clearly see that there is still a very small gap present between valence and conduction band just above the FL though it appeared to be gapless in Fig~\ref{Fig5}. As an earlier study by M.Marshall et al. \cite{Marshall} finds EMB to be a topological insulator, we have calculated the surface states on two different surfaces (001) and (100). We present the results in Fig~\ref{Fig7} (a) and (b) respectively. 

From Fig~\ref{Fig7} we observe that the surface states are gapped on (001) surface whereas the gap closes (100) surface. This feature of having gapped surface states along one surface and gap closing on the other surface characterizes the system as an AFM topological insulator \cite{sur}. The gap opening and closing of the surface states can be understood from the symmetry considerations as discussed below. EMB is centrosymmetric and thus has inversion symmetry (P). In the A-AFM configuration the TRS (T) is broken due to intrinsic magnetism. However, an effective TRS (defined by the combination of the T and a lattice translation by half of the lattice constant along c-axis (t$_{1/2}$) (note that in A-AFM the magnetic unit cell is doubled along c) known as nonsymmorphic TRS i.e, S=Tt$_{1/2}$) is preserved. When we calculate the surface states on (001) surface, we set an open boundary along $c$ direction. This breaks the S symmetry leading to a gap in the surface states. Whereas when we look at the (100) surface we set the open boundary along $a$ direction. This restores the S sysmmetry giving rise to gapless surface states as seen in Fig~\ref{Fig7}(b). 
%%%%%%%%%%%%%%%%%%%%%%%%%%%%%%%%%%%%%%%
\begin{figure}
	%\vspace{-1.0cm}
	\begin{center}
		\includegraphics[width=8.0cm]{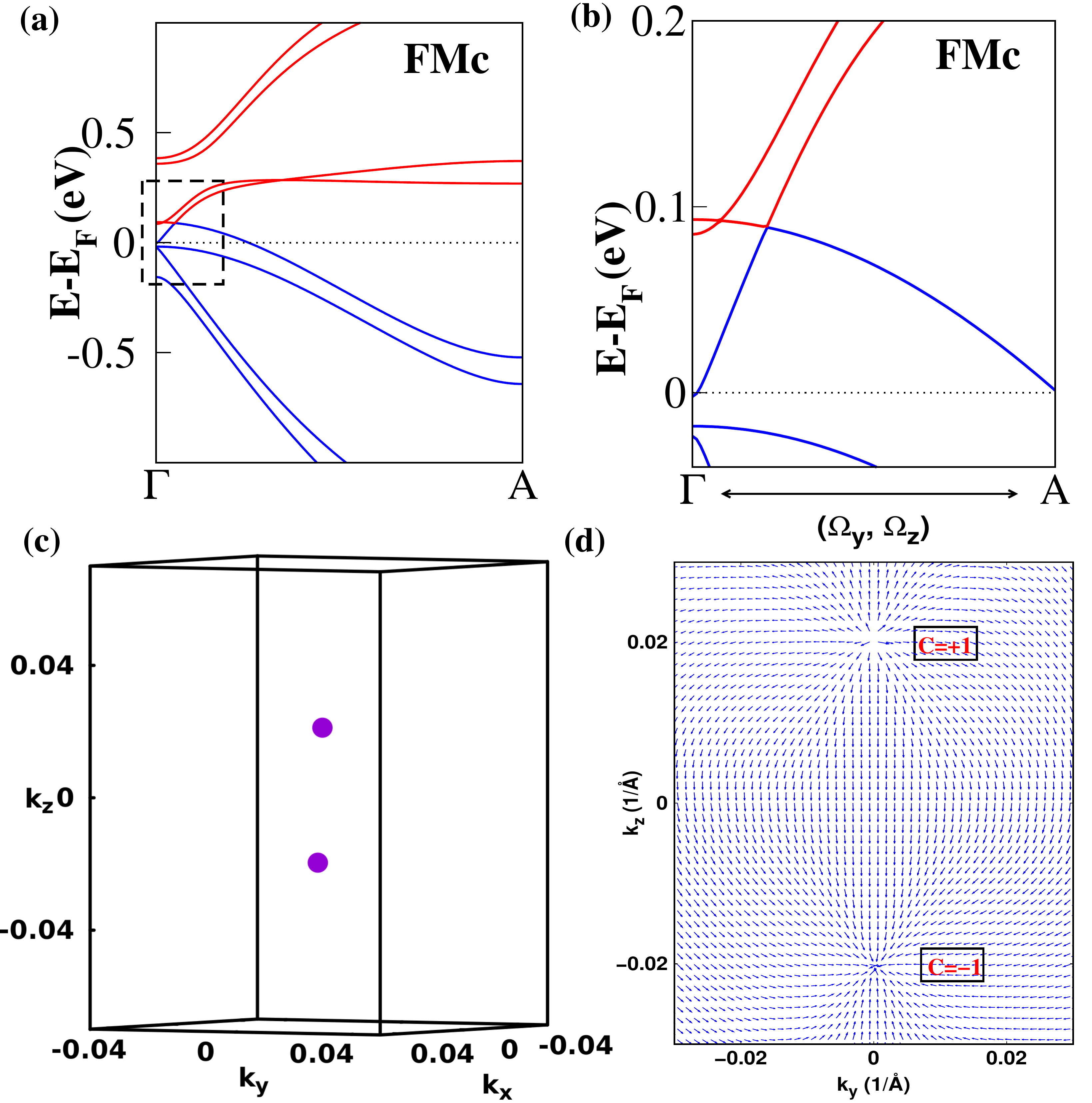}
	\end{center}	
	\caption{ (a) Bulk band structure for the EMB in FMc phase along $\Gamma$-A direction, (b) Zoomed in picture showing clearly the Weyl point, (c) The positions of Weyl points in 3D BZ, (d) Berry curvature on the k$_y$-k$_z$ plane showing source and sink at the pair of Weyl points.}
	\label{Fig8}
	
\end{figure}
%%%%%%%%%%  
%%%%%%%%%%%%%%%%%%%%%%
\begin{figure}
	%\vspace{-1.0cm}
	\begin{center}
		\includegraphics[width=8.0cm]{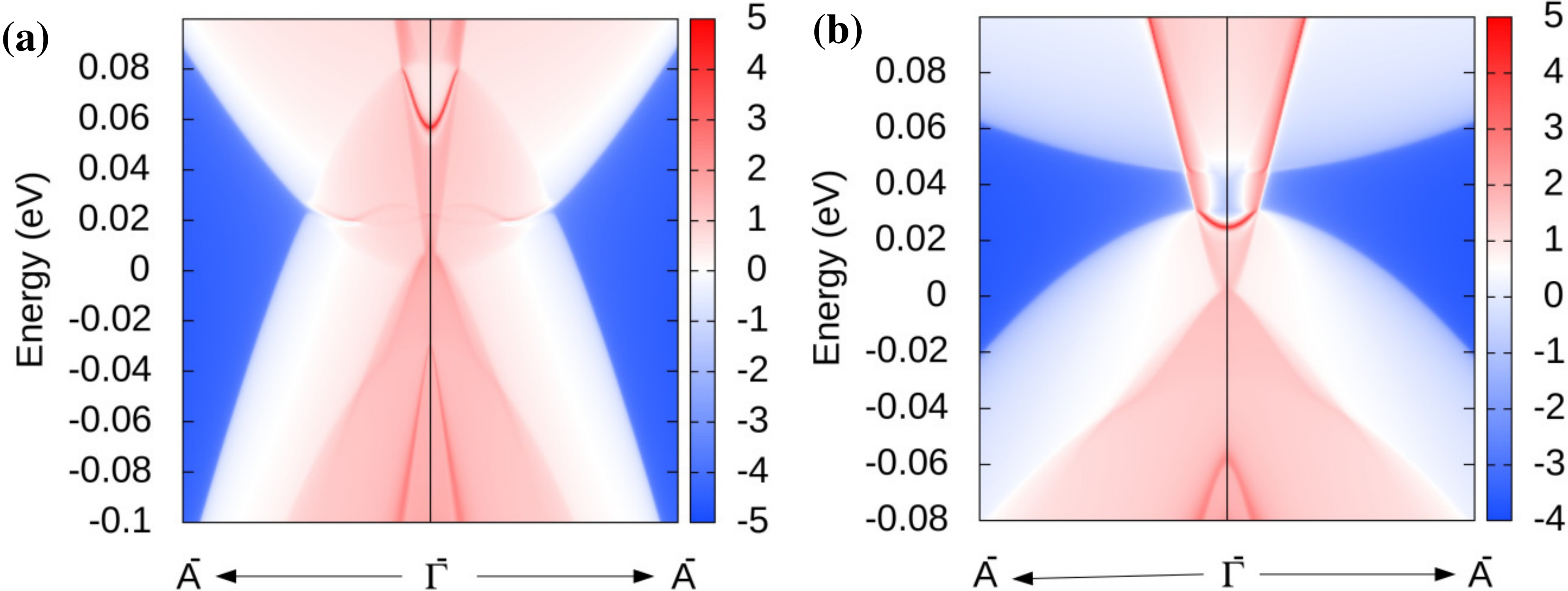}
		\caption{Calculated ARPES surface states for FMc configuration for (a) EMB (b) Eu$_{0.5}$Ca$_{0.5}$Mg$_2$Bi$_2$ on (100) surface.}
		\label{Fig9}
	\end{center}
\end{figure} 
%%%%%%%%%%%%%%%%%%%%%%%%%%%%%%%%%%%%%%%

Next we look at the same region of the band structure plot for A-AFMc configuration which is energetically closest to the ground state. We present the result in Fig~\ref{Fig6}(b). We observe that the tiny band gap that was present in A-AFMb case has become a band crossing point between valence and conduction band which are slightly away from $\Gamma$ point along A-$\Gamma$-A direction(inset of Fig~\ref{Fig6}(b)). These two points are four-fold degenerate Dirac points as we explain below. As discussed above the system possess both S and P symmetries in A-AFMc phase and it can be shown that the combined operation of S and P is antiunitary, i.e. (SP)$^2$=-1. Therefore, at  each ${\bf k}$ point in BZ the bands are doubly degenerate due to Kramer's theorem. Thus the band crossing point between valence and conduction bands is four-fold degenerate Dirac point. In addition to S and P, the A-AFMc magnetic configuration (where the Eu magnetic moments point along $c$-axis) posseses C$_{3z}$ rotational symmetry (which was not present in A-AFMb) which protects a stable pair of Dirac points as we observe in Fig~\ref{Fig6}(b)\cite{sur}. With the simultaneous presence of inversion (P), nonsymmorphic TRS (S) and threefold rotational symmetry (C$_{3z}$), EMB in its AFMc magnetic state can be categorized as class-I Dirac semimetal as per the classification proposed by Yang and Nagaosa \cite{nagaosa} where the Dirac points appear away from the high symmetry points. Thus we see that in A-AFM state by changing the direction of Eu magnetic moments from $b$ to $c$ direction, one can turn this system from a AFM topological insulator\cite{mong} in to a Dirac semimetal due to restoration of threefold rotational symmetry. By looking at the energy difference between A-AFMb and A-AFMc from Table~\ref{Tab1}, we see that it is only 0.2meV per formula unit which is very small and therefore can be accessible by external magnetic field. 

Finally, we make the most interesting observation when we explore the same $\Gamma$-A path in case of FM state with Eu moments along $c$ direction (i.e. FMc). The two fold spin degeneracy of the electronic bands are now lifted due to the breaking of S symmetry giving rise to Weyl points very close to the FL (see Fig~\ref{Fig8}(a) and (b)). Interestingly, in Fig~\ref{Fig8} (a) and also in the zoomed-in plot Fig~\ref{Fig8} (b) we observe only one crossing between valence and conduction band which means we have only a single pair of Weyl points here. Since we had two Dirac points in A-AFMc phase we expected to obtain four (two pairs of) Weyl points. To confirm whether we have a single pair and two pairs of Weyl points, we calculated the angle resolved photoemission spectroscopy (ARPES) spectrum of the surface states using WT which we present in Fig~\ref{Fig9}(a). One can clearly observe a single Fermi arc connecting the Weyl points which therefore confirms that we indeed have a single pair of Weyl points in EMB in its FMc phase similar to that observed in EuCd$_2$As$_2$. As the two Dirac points from which these Weyl points have emerged, were located very close to the $\Gamma$ point and the exchange splitting is quite large, we believe the other pair of Weyl points have annihilated each other exactly like the situation in EuCd$_2$As$_2$\cite{AA}. However, unlike EuCd$_2$As$_2$ the FMc state in EMB is not half-metallic without SOC\cite{AA} (see Fig~\ref{Fig4} (a)). Thus the FMc phase of EMB is a Weyl semimetal with a single pair of Weyl points. The positions of this pair of Weyl points, their energies and Chern numbers are calculated by WT and are listed in Table~\ref{Tab3}. The gap smaller than 10$^{-5}$ eV is considered to be gapless. The positions of these two Weyl points are shown in 3D BZ in Fig ~\ref{Fig7}(c). The corresponding Berry curvatures are also calculated and are shown in Fig ~\ref{Fig7}(d) cleary showing the source and sink of the spin texture. Since the total energy difference between the ground state A-AFMb and FMc configuration is about 1.39 meV per formula unit which is very small and FMc phase should be possible to achieve via application of a moderate external magnetic field. EuMg$_2$Bi$_2$ is therefore a very versatile compound where multiple topological phases can be seen depending on the magnetic order. One can go from topological insulating state to Dirac semimetallic state and then to Weyl semimetallic state by varying the magnetic order from A-AFMb to A-AFMc and then to FMc. As these magnetic states are energetically close to each other an external magnetic field can be used as a handle to tune magnetic order as well as the corresponding topological phase making it a suitable candidate for magnetically driven topological system. 

%%%%%%%%%%%%%%%%%%%%%%
\begin{figure}
	%\vspace{-1.0cm}
	\begin{center}
		\includegraphics[width=8.0cm]{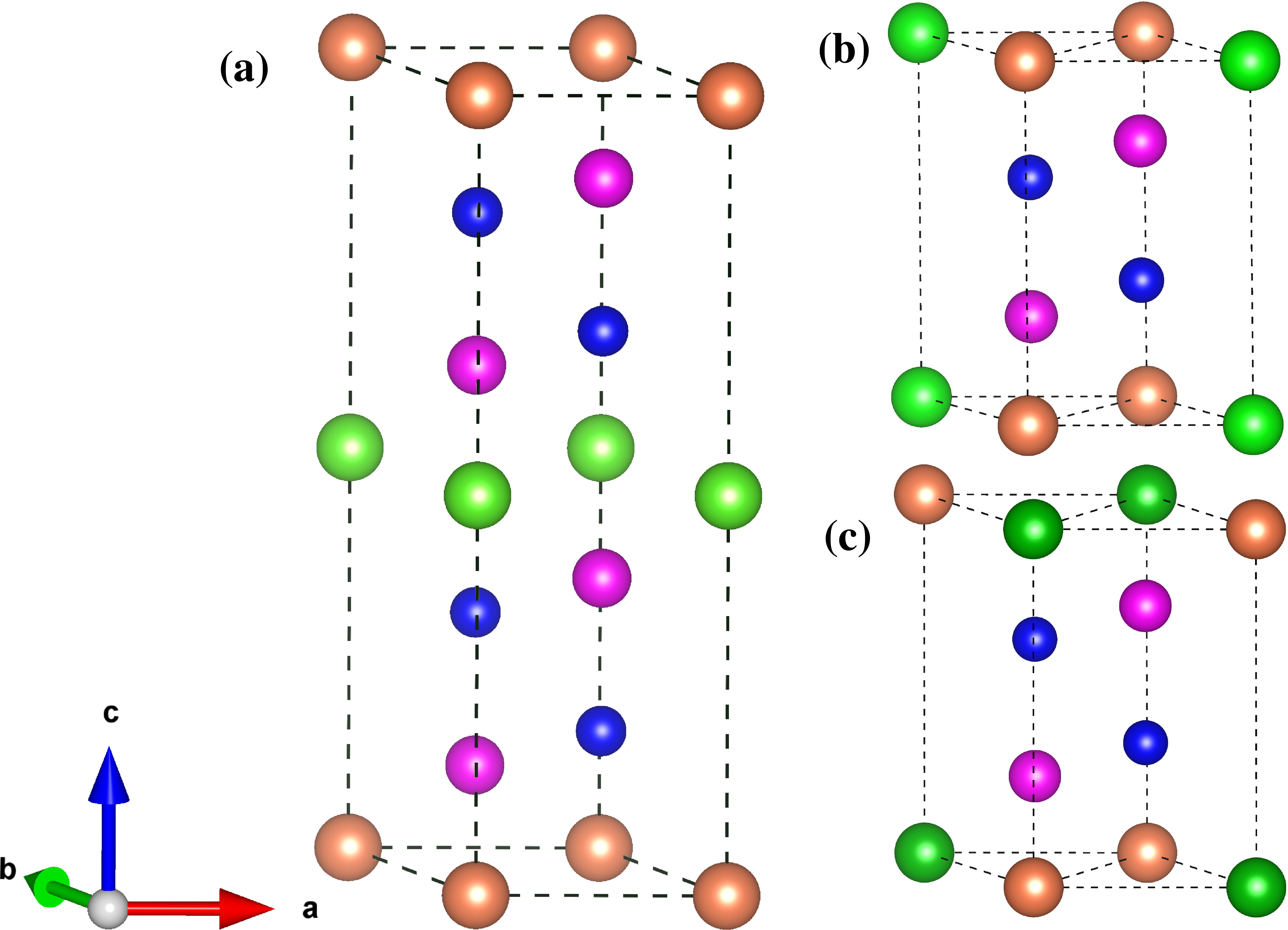}
		\caption{Supercells showing various Eu(brown)/Ca(green) ordering considered our calculation (see text).}
		\label{Fig10}
	\end{center}
\end{figure} 
     %%%%%%%%%%%%%%%%%%%%%%%%%%%%%%%%%%%%%%%
 
%%%%%%%%%%%%%%%%%%%%%%%
\subsection{Ca doped EuMg$_2$Bi$_2$: Topological properties}  
\begin{figure}[!htp]
	\includegraphics[width=9.0cm]{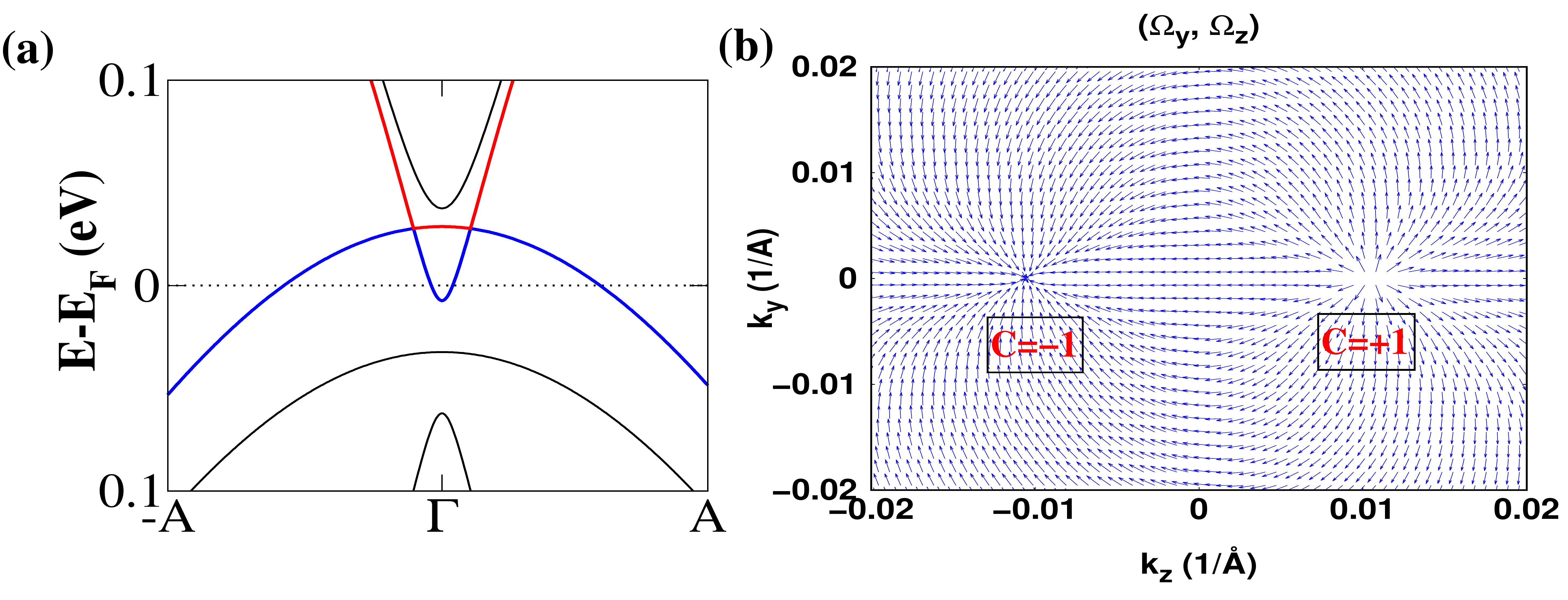}
	\caption{(a) Bulk band structure of Eu$_{0.5}$Ca$_{0.5}$Mg$_2$Bi$_2$ for FMc state along $\Gamma$-A direction (b) Berry curvature on the k$_y$-k$_z$ plane source and sink around Weyl points.  }
	\label{Fig11}
	% 	\end{center}
\end{figure}

In a recent experimental work, Marshall et al.\cite{Marshall} have reported 50\% and 67\%  doping of Ca at Eu sites in EuMg$_2$Bi$_2$ where the authors observed that the ground state magnetic order changes from A-AFM to FM upon 67\% Ca doping. In order to explore the effect of Ca doping on the magnetic order and also the topological properties which is seen to be strongly related to the magnetic order in the parent compound, we have performed density functional theory calculations for 50\% and 67\%  Ca doped EuMg$_2$Bi$_2$. The structural parameters for the 50\% doped compound is taken from the experimental paper\cite{Marshall}. We considered three types of Eu-Ca site ordering for half-doped compound. In one case we have the 2D layers of Eu and Ca alternately stacked along $c$-axis as shown in Fig~\ref{Fig10}(a), in the second case Eu and Ca ions alternate in each layer as shown in Fig~\ref{Fig10}(b)and in third case Eu and Ca alternate in all three directions as shown in Fig~\ref{Fig10}(c). After structural optimization we observe that the first one is lowest in energy. Hence we present below the results considering the supercell shown in Fig~\ref{Fig10}(a). Our total energy calculations within GGA+U+SO approximation shows that the ground state of Eu$_{0.5}$Ca$_{0.5}$Mg$_2$Bi$_2$ is still A-type antiferromagnet with Eu moments pointing along $b$ direction (A-AFMb) similar to the parent compound. However, the energy difference between FMc and A-AFMb is only about 0.04 meV per formula unit which is much smaller compared to the same in the parent compound. This is also consistent with the experimental report\cite{Marshall1} where the authors observed A-AFM magnetic state for 50\% Ca doping at Eu sites with a lower transition temperature. Looking into the electronic structure along $\Gamma$-A direction of FMc magnetic configuration, we observe very interesting topological feature as shown in Fig~\ref{Fig11}. One can clearly see in this case the presence of a single pair of Weyl points as well which are located even closer to the Fermi level. The calculated ARPES surface states Fig~\ref{Fig9}(b)  clearly shows the presence of a single Fermi arc joining the two Weyl points. Comparing to the parent compound (Fig~\ref{Fig8}(d)), one also observes that the distance between the pair of Weyl points decreases in the doped compound (i.e. points are getting closer to the $\Gamma$ point) than the corresponding distance in the parent compound. We have listed in Table~\ref{Tab3} the positions of this pair of Weyl points with respect to the FL, their energies and Chern numbers. In Fig ~\ref{Fig11}(b) we present the Berry curvature corresponding to the pair of Weyl points. 
    %%%%%%%%%%%%%%%%%%%%%%%%%%%%%%%%%%%%%%%

\begin{figure}
 	%\vspace{-1.0cm}
 	\begin{center}
 		\includegraphics[width=8.0cm]{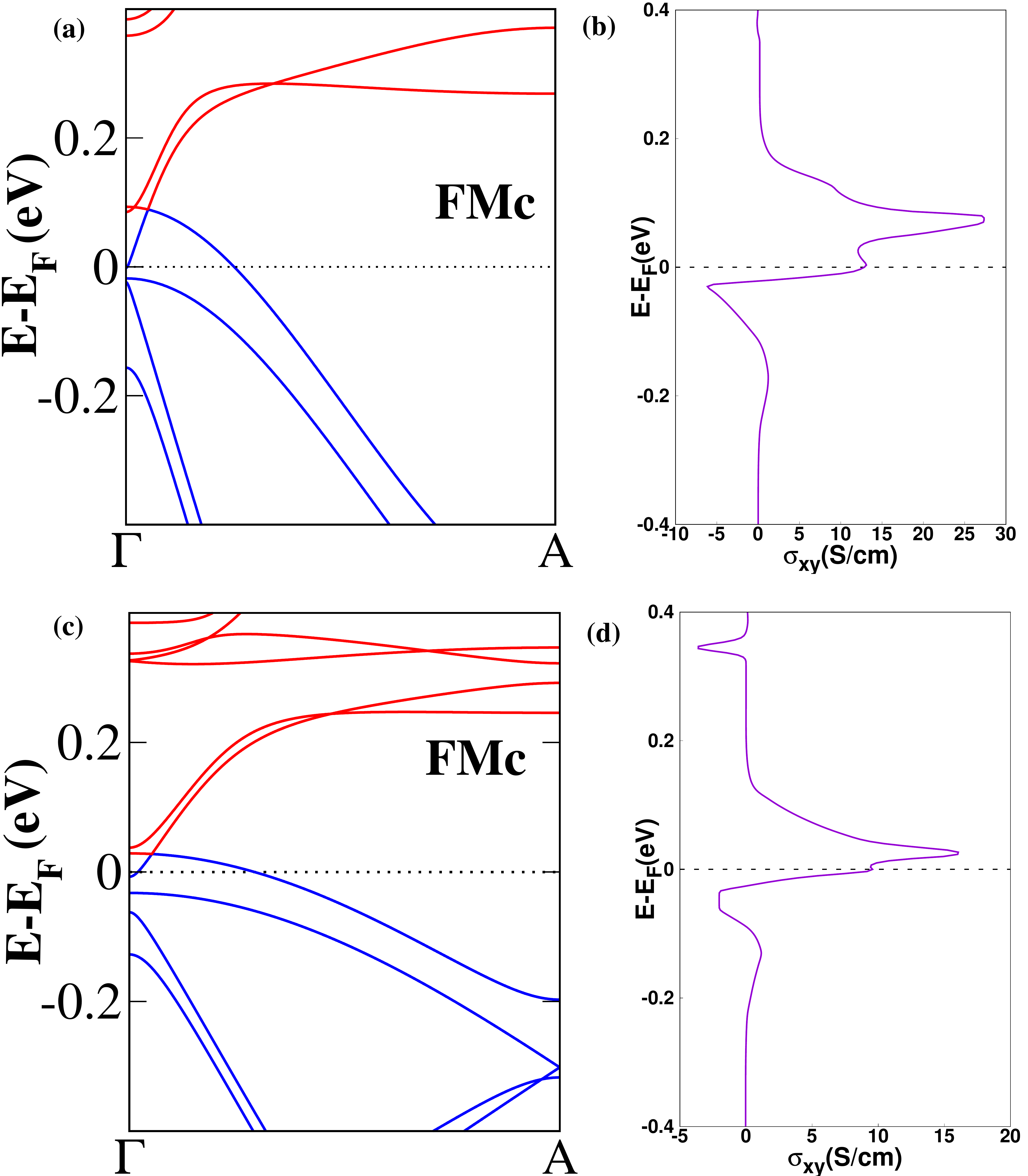}
 		\caption{ Band structure of FMc state along $\Gamma$-A direction along with corresponding AHC for (a) EuMg$_2$Bi$_2$ and (b)Eu$_{0.5}$Ca$_{0.5}$Mg$_2$Bi$_2$.}
 		\label{Fig12}
 	\end{center}
 \end{figure}

Further, the Berry curvature dependent intrinsic Anomalous Hall Conductivity (AHC) was computed by the Wannier function-based Hamiltonian, which is presented as a function of energy (E-E$_F$) in Fig ~\ref{Fig12} for parent as well as 50\% Ca doped compound alongside the corresponding band structures. The peak of the AHC is present just above the Fermi energy at 0.08 meV exactly at the position of nondegenerate band crossing point (Weyl Point) in case of parent compound (Fig ~\ref{Fig12} (b)) while the peak in the doped compound is situated at 0.027 meV above Fermi Level where the WP is located (Fig ~\ref{Fig12} (d)). However, the peak height of AHC in the doped compound is reduced compared to that in the parent compound. This is because the AHC depends on the separation between the Weyl points for a single pair of Weyl points\cite{AHC}. We have also performed the same calculation for 67\% Ca doped compound and observed that the distance between the Weyl points in this case further reduces and Weyl points move even closer to the FL implying that the AHC peak height will be further reduced in this case. Therefore, the Ca doping can be used as a handle to tune AHC in this compound.
  
 %%%%%%%%%%%%%%%%%%%%%%%%%%%%%%%%%%%%%%%%%%%%%%%%%%
 %%%%%%%%%%%%%%%%%%%%%%%%%%%%%%%%%%%%%%%

\section{Conclusions}
We performed a systematic and thorough investigation of the magnetic order, electronic structure, topological and related transport properties of EuMg$_2$Bi$_2$ and Ca doped EuMg$_2$Bi$_2$ using first principles density functional theory calculations. Our total energy calculations for various magnetic orders in EuMg$_2$Bi$_2$ reveal that A-type AFM with Eu magnetic moments pointing along crystallographic $b$ axis (A-AFMb) is the ground state. However, the A-AFM and FM states with Eu moments pointing along $c$ (A-AFMc and FMc respectively) are energetically very close exactly similar to the situation in another isostructural compound EuCd$_2$As$_2$ implying that an external magnetic field can be used to tune the magnetic state of the compound. Our detailed electronic structure calculations and topologocal property study further reveals that while EMB is a topological insulator in its A-AFMb state consistent with the previous reports, it turns into a Dirac semimetal in its A-AFMc state and to a Weyl semimetal in its FMc state. Therefore, the toplogical phase of the system is driven by the underlying magnetic order which can be tuned by an external magnetic field making it a very interesting candidate material for device applications. We also observed that in the Weyl semimetallic state this system hosts a pair of Weyl points very close to the Fermi level. Upon doping 50\% Ca we observe that the ground state magnetic order still remains A-AFMb but the energy difference between A-AFMb and FMc reduces a lot compared to that in the parent compound. Most interestingly, the FMc state in EMB hosts a single pair of Weyl points very close to the FL (0.081 eV) and with 50\% Ca doping at Eu sites the pair of Weyl points move even closer to FL (0.027 eV) which is highly desirable for application purposes. Our calculations of AHC as a function of energy in EMB and its Ca doped variants further show that the AHC peak positions and heights vary as we doped EMB with Ca. This can be understood by the fact that AHC peak height depends linearly on the separation between the pair of Weyl points. As the separation decreases with Ca doping, the AHC peak height also correspondingly decreases. This imply that chemical substitution at Eu site in this compound can open an avenue for tuning the transport properties such as AHC.

\section{Acknowledgements} 

TM acknowledges Science \& Engineering Research Board (SERB), India for funding through SERB-POWER research grant (no. SPG/2021/000443). AC acknowledges computational facility Param Ganga at IIT Roorkee and also Ministry of Education, India for fellowship. 
\label{conc}

\end{document}